\documentstyle[12pt]{article}

\newcommand{\be}{\begin{equation}}
\newcommand{\ee}{\end{equation}}
\newcommand{\osi}{$^{16}O$}
\newcommand{\caf}{$^{40}Ca$}

\begin{document}
\pagestyle{empty}
\title{Long-Range Correlations \\ in Closed Shell Nuclei}
\author{H.~M\"{u}ther
\\\\
Institut f\"{u}r Theoretische Physik\\ Universit\"{a}t T\"{u}bingen\\
D-72076 T\"{u}bingen, Germany
\\\\
and
\\\\
L.D.~Skouras
\\\\
Institute of Nuclear Physics\\
N.R.C.P.S. Demokritos \\Aghia Pa\-ras\-kevi GR 15310, Greece}
\maketitle

\date{\today}

\begin{abstract}
The effects of correlations on the bulk properties of nuclei are
investigated in large model spaces including up to 21 single-particle orbits.
The evaluation of the single-particle Green function is made feasible by means
of the BAGEL approximation. The spectral function for single-nucleon pick-up
and removal is investigated for the nuclei \osi\ and \caf .
Special attention is paid to the effects produced by correlations on the
calculated ground state properties of closed shell nuclei.
It is observed that correlations beyond the
Brueckner Hartree Fock approximation tend to improve the results
obtained using  realistic nucleon nucleon interactions.
\end{abstract}
\pagestyle{empty}


\clearpage
\pagestyle{headings}

\section{Introduction}

The nuclear shell-model, which describes the nucleus as a system of
nucleons moving without correlations in a mean field, has been very
successful in explaining a large number of basic nuclear phenomena at
low energies. Nevertheless, it has become quite obvious that
to derive nuclear properties from a realistic nucleon-nucleon
(NN) interaction  it is necessary to consider the correlations between the
nucleons. This necessity arises from the fact that NN interactions,
which are adjusted to describe NN scattering data,
contain very  strong short-range components. As a consequence,
Hartree-Fock (HF) calculations employing such interactions predict bulk
properties of nuclei which are far off the empirical data and, in fact,
fail to account for the binding of nuclei.

Several methods have been developed to deal with the strong short-range
components of the NN interaction. One of these methods is to consider
that the wavefunctions contain short-range correlations, e.g. Jastrow
correlation functions, which are  optimized in a variational calculation
\cite{pand}.
An alternative approach is to introduce effective operators.
A typical example of such an effective operator
is the Brueckner G-matrix \cite{brueck,day}. This G-matrix is evaluated
from the NN interaction by solving the Bethe-Goldstone equation, which
accounts for the virtual excitation of two interacting nucleons into
single-particle states above the Fermi energy, i.e states unoccupied
by other nucleons. In solving the Bethe-Goldstone equation for realistic
NN interactions one has to consider 2-particle configurations up to an
excitation energy of a few GeV \cite{rev,polls,bonats}.

The consideration of the short range, or high-energy, correlations
in the Brueckner-Hartree-Fock (BHF) approach results in  a major
improvement on the prediction of nuclear properties, as compared to the
independent-particle or HF approach.
However, trying to evaluate the binding-energy and saturation density
of nuclear matter, or the binding-energies and radii of finite nuclei
within the BHF approximation, one still observes significant deviations
from the corresponding data \cite{mu1}. Motivated by the success of the
Walecka model \cite{serot}, attempts have been made to include relativistic
effects within the so-called Dirac-BHF approach, leading to quite an
improvement for the description of nuclear matter \cite{brom}. The
corresponding calculation for finite nuclei, however, did not yield a
satisfying agreement between theory and experimental data \cite{fritz}.

It is worth noting that quite different results are obtained in the
studies of infinite nuclear matter and finite nuclei. This demonstrates
that the structure of finite nuclei is more complicated and cannot be treated
simply as a piece of nuclear matter of finite size.
These differences cannot be attributed to the short-range correlations, as the
configurations at high energies should be similar in both cases. It is,
therefore, more likely that the different behaviour of nuclear matter and
finite nuclei is due to the excitation modes  of these systems
at lower energies. At
such energies the excitation spectrum for a finite system still
reflects the shell structure, which is absent for the infinite system.

In this study, we investigate improvements to the BHF approach caused by a
detailed treatment of low- and medium-energy excitations. Specifically,
we investigate the effects produced  by  configurations with excitation energy
up to about 100 MeV. These  configurations are formed  from single-particle
states which all belong to a few major shells around the Fermi level.
The effects produced on the nuclear wavefunction by the admixtures of these
configurations will be described in the following as long-range correlations,
in contrast to the short-range correlations discussed above.

For the study of these long-range correlations we  employ a
Green-function approach within a finite model space of discrete
single-particle states \cite{rev,brand,neck,sko1}.
We describe the irreducible self-energy for the nucleons in terms of the HF
contribution plus terms of higher order in the residual NN interaction,
which are described by the diagrams shown in Fig. 1.
For the residual interaction ${\cal G}$ we consider a G-matrix, which sums up
all contributions originating from high-energy 2-particle excitations,
i.e excitations  outside the space of our calculation (model-space).
Thus all intermediate states displayed  in Fig. 1 must be understood to belong
to the model-space under consideration. Our definition of the nucleon
self-energy is different from that employed in the BHF approximation,
which includes diagram 1a, and those containing particle-particle ladder
diagrams like 1c, but ignores contributions like those described by
diagrams 1b, 1d and 1e.  This non-symmetric treatment of
particle-particle and hole-hole excitation is a typical feature of the
Brueckner hole-line expansion. It may very well be justified for the
study of short-range correlations, as the high-energy particle-particle
configurations do not have any counterpart on the hole-hole side. The
ordering of the hole-line expansion, however, may be questionable if
low-energy excitations are to be considered. Therefore we would like to
consider in our approximation for the self-energy, the contribution of
all terms with intermediate 2 particle-1 hole and 2 hole-1 particle
states (see Fig. 1) including all interactions between those states.

For the irreducible self-energy one may then solve the Dyson equation
to determine the reducible self-energy or the corresponding
single-particle Green-function. In the Lehmann representation this
Green function is defined in terms of many poles below and above the
Fermi-energy and the value of the spectral function at these poles. In
ref.~\cite{sko1} the so-called BAGEL approach has been introduced to represent
the complicated structure of the Green-function in terms of a few
``characteristic'' poles. Using this approximate representation,
it becomes possible to
evaluate a self-consistent Green-function in the sense that the irreducible
self-energy is calculated using the Green-function, which, in turn,
is obtained from the Dyson equation for this self-energy \cite{sko2}.

One of the aims of the present study is to investigate  the
sensitivity of the results for the Green-function or the spectral function on
the approximations chosen for the irreducible self-energy and on the choice
of the model-space. As examples we  consider the closed shell nuclei
\osi and \caf . The results obtained for the spectral function can
directly be compared with experimental data obtained in $(e,e')p$
experiments \cite{witt,grab,kramer}. From the Green-function we can
furthermore derive the occupation probabilities for the various
single-particle orbits. These occupation probabilities are quite a
conspicuous measure for the deviation from the independent particle
model, i.e. a measure for the amount of correlation effects. Such
occupation probabilities may be compared to results obtained for
nuclear matter by using the Green-function approach \cite{rev,vonder},
within Brueckner theory \cite{mah,koehler} or by means of the correlated
basis function theory \cite{benh}.

Finally, one may also use the Green-function  to determine the
ground-state energy and other observables like the radius. One can
study the effects of correlations beyond the BHF approximation and
estimate the importance of particle-particle and hole-hole
ladder diagrams on the saturation properties of infinite nuclear
matter \cite{kuo1,kuo2}.

The Green-function approach is reviewed briefly in sect. 2.
The presentation there  helps to establish the nomenclature but also to
define some details of the present investigation. Results on the
spectral function, occupation probabilities, binding energies and radii
for \osi\ and \caf\, obtained with the use of various approximations, are
presented in sect. 3. The main conclusions are summarized in sect. 4.

\section{BAGEL approximation for the one-body Green-function}

\subsection{Model Space and effective Hamiltonian}

As already mentioned in the introduction, one of the main aims of the present
investigation is to explore effects of long-range correlations by means of the
Green-function approach within a certain truncated model-space.
On the other hand, the effects of short-range correlations
are taken into account by the introduction of an effective interaction,
i.e  a G-matrix appropriate for the model-space.
Such a concept of a double-partitioned Hilbert space has been used before for
finite nuclei \cite{sko1,bar} and infinite nuclear matter \cite{kuo2,kuo3}.
The G-matrix is determined as  the solution of the Bethe-Goldstone equation
\be
{\cal G} = V + V \frac{Q_{\hbox{mod}}}{ \omega - Q_{\hbox{mod}} T
Q_{\hbox{mod}}} {\cal G}\; .
\label{eq:betheg}
\ee
In this equation $T$ is the kinetic energy  operator while
$V$ stands for the bare two-body interaction. For the latter we have chosen
the One-Boson-Exchange potential $C$ defined in table A.1 of \cite{rupr}.
The Pauli operator $Q_{\hbox{mod}}$ is defined in terms of harmonic-oscillator
single-particle states. Thus applying $Q_{\hbox{mod}}$ to two-particle states
$\vert \alpha\beta >$ one obtains
\be
 Q_{\hbox{mod}} \vert \alpha\beta > = \left\{ \begin{array}{ll}
0 & \mbox{if $\alpha$ or $\beta$ below Fermi level}\cr
0 & \mbox{if $\alpha$ and $\beta$ in model space} \cr
\vert \alpha\beta > & \mbox{else} \end{array} \right.
\label{eq:qmod}
\ee
Here the Fermi level refers to the Fermi level of the independent
particle model for the nucleus under consideration.  The model-space in
eq.~(\ref{eq:betheg}) consists of all single-particle states up to
and including the 0h1f2p shell of the oscillator potential, while a constant
value of $\omega = -30$ MeV has been adopted for the starting energy.
This choice of using a constant starting energy in
the G-matrix is of course an approximation introduced to simplify the
calculations. In sect. 3 we discuss the sensitivity of our results
on the choice of  the starting energy. There we will also consider the
effects of using different model spaces.

A major problem in the use of the modern OBE versions of realistic NN
interactions, which are defined e.g. in \cite{rupr}, is the fact that BHF
calculations using these interactions yield predictions for the radii which
are too small  compared to the experimental data \cite{carlo}.
This has serious consequences on calculations beyond the mean field
approximation. If one uses e.g.~such BHF wave functions to evaluate the
residual interaction to be used in shell-model calculations, one obtains
results which are in poor agreement with the experimental data. On the
other hand, evaluating the matrix elements for a set of harmonic oscillator
wave functions which yield accurate radii, leads to matrix elements for
the shell-model calculation, which provide a fair agreement with
experiment \cite{mort}.
Therefore one is tempted to modify the hamiltonian in such a way that
a HF calculation using $\cal G$ yields a self-consistent basis of oscillator
wave functions which reproduce the experimental radius. This is achieved
by subtracting from the $\cal G$ defined in eq.(\ref{eq:betheg})
the non-diagonal terms of the HF single-particle hamiltonian in the
oscillator basis
\be
\Delta {\cal G} = \sum_{\alpha \ne \beta} \left( T_{\alpha\beta}
+ U_{\alpha\beta} \right) a^\dagger_\alpha a_\beta \; ,
\label{eq:adjust}
\ee
where $T_{\alpha\beta}$ and $U_{\alpha\beta}$ define the matrix elements
of the kinetic energy and the HF single-particle potential , respectively,
calculated in a harmonic oscillator basis.  This basis is
defined by an oscillator parameter of $b=$ 1.76 fm for all studies of
\osi\ and $b=$ 2.0 fm in case of the nucleus \caf .

\subsection{Green function and matrix diagonalisation}

As a first step towards the evaluation of the single-particle Green
function we determine the contributions to the irreducible self-energy,
which are of second order in the residual interaction:
\be
\Sigma^{(2)}_{\alpha\beta}(\omega ) =  \frac{1}{2}  \sum_{\gamma
\delta \mu} \int \frac{d \omega_1}{2 \pi i} \int \frac{d \omega_2}{ 2 \pi i}
< \alpha \mu \vert {\cal G} \vert \gamma \delta> < \gamma
\delta \vert {\cal G} \vert \beta \mu >
g_\gamma (\omega -
\omega_1 + \omega_2 ) g_\delta (\omega_1 ) g_\mu (\omega_2)
\; . \label{eq:self2}
\ee
Note that the summation on single-particle states $\gamma \, , \,  \delta \,
, \, \mu$ is restricted to states within the model-space.
For discrete states $\alpha$, the
$g_{\alpha}(\omega )$ denote the Hartree-Fock (HF) approximation
for the Green function  \cite{sko1}.
This equation can be rewritten into the form
\be
\Sigma^{(2)}_{\alpha\beta}(\omega ) =  \frac{1}{2}  \sum_{p_{1}, p_{2}, h}
\frac{< \alpha h \vert{\cal G} \vert p_{1} p_{2}> < p_{1} p_{2} \vert
{\cal G} \vert \beta h >}{\omega - e(p_{1}, p_{2}, h) +i\eta}
\; + \; \frac{1}{2}  \sum_{h_{1}, h_{2}, p}
\frac{< \alpha p \vert{\cal G} \vert h_{1} h_{2}> < h_{1} h_{2} \vert
{\cal G} \vert \beta p >}{\omega - e(h_{1}, h_{2}, p) -i\eta }
\label{eq:self2b}
\ee
where we have introduced the abbreviation
\be
e(\alpha, \beta, \gamma ) = \epsilon^{HF}_\alpha + \epsilon^{HF}_\beta
- \epsilon^{HF}_\gamma
\label{eq:2p1hen}
\ee
and the $\epsilon^{HF}_\alpha$ are the HF single-particle energies. In
eq.~(\ref{eq:self2b}) the summations on particle labels like $p_{1},
p_{2}$ and $p$ are restricted to those single-particle states within the
model space, which are above the Fermi level, whereas the labels
$h_{1}, h_{2}$ and $h$ refer to hole states. It is evident that the
first term on the right hand side of eq.~(\ref{eq:self2b}) refers to the
2 particle - 1 hole  contribution to the
self-energy (diagram a of Fig.~1) while the second term defines
2 hole - 1 particle contribution (diagram 1b).
This self-energy can now be inserted into a Dyson equation for
the Green function $G_{\alpha \beta} (\omega )$ taking into
account the correlation effects contained in $\Sigma^{(2)}$
\begin{eqnarray}
G_{\alpha \beta } (\omega ) &= \delta_{\alpha \beta}
g_{\alpha} (\omega ) + \sum_{\gamma } g_{\alpha } (\omega ) \Sigma_{\alpha
\gamma } (\omega ) G_{\gamma \beta } (\omega ) \label{eq:dyson1}\\
&= \sum_n \frac{<\Psi_0^A \vert a_{\alpha} \vert \Psi_n^{A+1} >< \Psi_n^{A+1}
\vert a^\dagger_\beta \vert \Psi_0 >}{ \omega -\omega_{n}^+ +
i\eta }\nonumber\\ & \qquad +
\sum_m {<\Psi_0^A \vert a_{\beta}^\dagger \vert \Psi_m^{A-1} >< \Psi_m^{A-1}
\vert a_\alpha \vert \Psi_0 > \over \omega - \omega_{m}^- - i\eta }
\; . \label{eq:dyson2}
\end{eqnarray}
The second line (eq.(\ref{eq:dyson2})) exhibits the Lehmann representation
of the single-particle Green function in terms of the spectroscopic
amplitudes $<\Psi_0^A \vert a_{\alpha} \vert \Psi_n^{A+1} >$ and\\
$< \Psi_n^{A+1}\vert a^\dagger_\alpha \vert \Psi_0 >$,
where
$a_\alpha$ ($a_\beta^\dagger$) stands for the single-particle annihilation
(creation) operator in the HF basis. The state $\Psi_0^A$ refers to the
ground-state of the A-particle system, while $\Psi_n^{A+1}$ ($\Psi_m^{A-1}$)
stands for the states of the A+1-particle (A-1) system as obtained in the
present approach.
Also the energy variables
\begin{eqnarray}
\omega_n^+ &= E_n^{A+1} - E_0^A \; ,\\
\omega_m^- &= E_0^A - E_m^{A-1}   \; ,
\end{eqnarray}
are defined in terms of energies obtained for the states of the nuclei
with $A$, $A+1$ and $A-1$ nucleons. The substantial ingredients of the
Lehmann representation of eq.~(\ref{eq:dyson2}) can be obtained from a
solution of an eigenvalue problem \cite{sko1}
\be
\pmatrix{\epsilon_{\alpha}^{HF} & 0 & a_1 & \ldots & a_K & A_1 & \ldots &
A_L
\cr 0 & \epsilon_{\beta}^{HF} & b_{1}  & \ldots & b_{K}  & B_{1} &
\ldots & B_{L} \cr
a_1 & b_{1} & e_1 & & & 0 & & \cr \vdots & \vdots & & \ddots & & &  & \cr
a_K & b_{K} & 0 & & e_K & & & 0 \cr A_1 & B_{1} &  & & &  E_1 & & \cr
\vdots &\vdots& & & &
& \ddots & \cr A_L & B_{L}& 0 & \ldots & 0 & & \ldots & E_L \cr }
\pmatrix{ X_{0,\alpha}^n \cr X_{0,\beta}^n \cr X_1^N \cr \vdots
\cr X_K^n \cr Y_1^n \cr \vdots \cr Y_L^n
\cr }
= \omega_n
\pmatrix{ X_{0,\alpha}^n \cr X_{0,\beta}^n\cr X_1^n \cr \vdots \cr
X_K^n \cr Y_1^n \cr \vdots \cr Y_L^n
\cr } \; .
\label{eq:matr1}
\ee
In writing this equation, we assume that for a given set of conserved
quantum numbers (parity, isospin, angular momentum) we have in our
model space two HF single-particle states ($\alpha$, $\beta$) and $K$ 2p1h
configurations  with energies (see definition of eq.~(\ref{eq:2p1hen}))
\be
e_{i} = e (p_{1}, p_{2}, h) \; ,
\ee
and connecting matrix elements
\begin{eqnarray}
a_{i} & = < \alpha h \vert{\cal G} \vert p_{1} p_{2}> \nonumber\\
b_{i} & = < \beta h \vert{\cal G} \vert p_{1} p_{2}>\; .
\end{eqnarray}
Furthermore we consider $L$ 2h1p configurations of the same symmetry
as the single-particle states. These have energies
\be
E_{j} = e (h_{1}, h_{2}, p) \; ,
\ee
and connecting matrix elements
\begin{eqnarray}
A_{j} & = < \alpha p \vert{\cal G} \vert h_{1} h_{2}> \nonumber\\
B_{j} & = < \beta p \vert{\cal G} \vert h_{1} h_{2}>\; .
\end{eqnarray}
Solving eq.~(\ref{eq:matr1}) one obtains eigenvalues $\omega_{n}$. These
eigenvalues must be identified with $\omega_{n}^+$ of the Lehmann
representation of eq.~(\ref{eq:dyson2}), if $\omega_{n}$ is an energy
above the Fermi energy\footnote{The definition of the Fermi energy is given
below}. In this case we have
\be
<\Psi_0^A \vert a_{\alpha} \vert \Psi_n^{A+1} > =  X_{0,\alpha}^n \; .
\ee
If $\omega_{n}$ is below the Fermi energy we obtain
$\omega_{n}=\omega_{n}^-$ and
\be
<\Psi_0^A \vert a_{\alpha}^\dagger \vert \Psi_n^{A-1} > =  X_{0,\alpha}^n
\; .
\ee

In a straightforward way one can improve the approximation discussed so
far and incorporate the effects of residual interactions between the
2p1h and 2h1p configurations, i.e.~account for diagrams like those
displayed in Fig. 1c, 1d and 1e in the definition of the self-energy.
One simply has to modify the corresponding parts of the matrix in
eq.(\ref{eq:matr1}) and replace
\be
\pmatrix{e_{1} & \ldots & 0 \cr \vdots & \ddots &\cr 0 &\ldots &e_{K}\cr}
\Longrightarrow {\cal H}_{2p1h} \; ,\qquad\mbox{and}\qquad
\pmatrix{E_{1} & \ldots & 0 \cr \vdots & \ddots &\cr 0 &\ldots &E_{K}\cr}
\Longrightarrow {\cal H}_{2h1p} \; ,
\label{eq:resid}
\ee
where ${\cal H}_{2p1h}$ and ${\cal H}_{2h1p}$ contain the
residual interactions in the 2p1h and 2h1p sub-spaces.

The normalization of the eigenvectors of eq.~(\ref{eq:matr1}) ensures
that
\begin{eqnarray}
\sum_{n} \vert X_{0,\alpha}^n \vert ^2 \quad + \quad \sum_{m} \vert
X_{0,\alpha}^m \vert ^2 & \quad = \nonumber \\
\sum_n \vert <\Psi_n^{A+1} \vert a^\dagger_\alpha \vert \Psi_0^A > \vert^2
\quad + \quad \sum_m \vert <\Psi_m^{A-1} \vert a_\alpha \vert \Psi_0^A >
\vert^2 & \quad = \quad 1 \; , \label{eq:norm}
\end{eqnarray}
where the sum on $n$ accounts for all solutions with
$\omega_n^+$ larger than the Fermi energy and the sum on $m$ for all
solutions with energies $\omega_m^-$ below the Fermi energy. These
spectroscopic factors can also be used to determine the occupation
probabilities
\be
N_{\alpha} = \sum_m \vert <\Psi_m^{A-1} \vert a_\alpha \vert \Psi_0 >
\vert^2 \; .
\label{eq:occu}
\ee
The approach discussed up to now, however, is not number-conserving in the
sense that it is not guaranteed that
\be
\sum_{\alpha } N_{\alpha} (2j_\alpha +1) = A \; ,
\label{eq:numb}
\ee
with  $(2j_\alpha +1)$ representing the degeneracy of the states $\alpha$.
This equation, however, allows one to determine the Fermi energy in such a
way that the particle number calculated according to the left hand side of
eq.~(\ref{eq:numb}) is as close as possible to the mass number $A$ of the
nucleus under consideration.
 An approach
which is strictly number-conserving is obtained only if the Green
function used to calculate the self-energy (like in eq.~(\ref{eq:self2}))
is consistent with the Green function resulting from the solution of
the Dyson equation (\ref{eq:dyson1}) \cite{baym,sko2}.

With the help of the single-particle Green function $G_{\alpha\beta}$
one can evaluate the expectation value for an arbitrary single-particle
operator $\hat O$
\be
< \Psi_0 \vert \hat O \vert \Psi_0 >  = \int_C {d \omega \over 2 \pi i}
\sum_{\alpha \beta} <\alpha \vert \hat O \vert \beta > G_{\alpha\beta}
(\omega ) \; ,
\label{eq:observ1}
\ee
where the $C$ below the integral sign denotes a contour integration
counter-clockwise in the upper half plane including the real axis.
Therefore only the contributions from the poles at energies below the
Fermi energy have to be considered and using the nomenclature of the
matrix representation in eq.~(\ref{eq:matr1}) this expectation value can
be rewritten as
\be
< \Psi_0 \vert \hat O \vert \Psi_0 > = \sum_{\alpha \beta m }
<\alpha \vert \hat O \vert \beta > X_{0,\alpha}^m X_{0,\beta}^m \; ,
\ee
where the sum on $m$ is restricted to solutions with energies
$\omega_{m}^-$ below the Fermi energy. The Green
function can furthermore be used to evaluate the binding energy of the
A-nucleon system by
\be
E_0^A = {1 \over 2} \int_C {d \omega \over 2 \pi i} \sum_{\alpha \beta}
\left[ <\alpha \vert T \vert \beta > + \delta_{\alpha\beta} \omega
\right]
G_{\alpha\beta} (\omega )\; , \label{eq:energy}
\ee
where $T$ denotes the kinetic energy.

\subsection{The BAGEL approximation}

After we have rewritten the solution of the Dyson equation in terms of an
eigenvalue problem (eq.~(\ref{eq:matr1})), it is now straightforward to
introduce the BAGEL approximation. The BAGEL approximation has
originally been formulated \cite{taigel} to define an appropriate
truncation scheme for shell-model calculations in large model spaces.
Since then this method has been employed in several calculations
\cite{leitner}, including
attempts to derive effective operators for truncated model spaces
\cite{leo}. The manner in which the BAGEL approximation can be used
to determine the single-particle Green function
has already been discussed in \cite{sko1,sko2}, so that only a brief
outline of the method need to be given here.

For many applications of the single-particle
Green function it is not necessary to know in detail all of its poles in the
Lehmann representation. It
should rather be sufficient to describe the energy distribution of the
spectroscopic amplitudes in terms of a few ``characteristic'' states of
the $A \pm 1$ systems.
For that purpose we consider the operator $\hat a$ which corresponds to
a part of the matrix in eqs.~(\ref{eq:matr1}) and (\ref{eq:resid})
\be
\hat a =
\pmatrix{\epsilon^{HF}_\alpha & a_1  &\ldots & a_K \cr
a_1 &  & &  \cr \vdots &  & {\cal H}_{2p1h} &  \cr a_K &  &  & \cr} \; ,
\label{eq:bag1}
\ee
and apply this operator on the single-particle state $\vert \alpha >$, which
in terms of the matrix representation of eq.~(\ref{eq:bag1}) is described
by the column vector $ (1, 0 \dots 0)^T$
\be
\hat a \vert \alpha > = \epsilon^{HF}_{\alpha} \vert \alpha > + \tilde
a_1 \vert \alpha_1 > \, ,
\ee
where $\vert \alpha_1 >$ is orthogonal to $\vert \alpha >$ and the coefficient
$\tilde a_1$ is chosen so that $\vert \alpha_1 >$ is  normalized.
Following the Lanczos algorithm \cite{lanczos}, one can subsequently
construct additional states $\vert \alpha_{i} >$, which are all
orthogonal to each other. Applying the Lanczos procedure $N$ times one
obtains $N$ basis states of the 2p1h configuration space, which have the
same symmetry quantum numbers as the single-particle state $\vert \alpha >$.
If now the  model-space contains another single-particle state
$\vert \beta >$  of the
same symmetry, one may consider in an analogous way the operator $\hat
b$ defined by
\be
\hat b =
\pmatrix{\epsilon^{HF}_\beta & b_1  &\ldots & b_K \cr
b_1 &  & &  \cr \vdots &  & {\cal H}_{2p1h} &  \cr b_K &  &  & \cr} \; ,
\label{eq:bag2}
\ee
and obtain by applying it initially to $\vert \beta >$
additional $N$ basis states $\vert \beta_{i}>$. In this case special care
should be taken that the states $\vert \beta_{i}>$ are not orthonormalized
only among themselves but also with respect to the $\vert \alpha_{i}>$.

In a similar
way we can furthermore construct for each single-particle state
$\vert \alpha >$,  $M$ basis states of the 2h1p configuration space by
considering the corresponding sub matrices of eq.(\ref{eq:matr1})
\be
\hat A =
\pmatrix{\epsilon^{HF}_\alpha & A_1  &\ldots & A_L \cr
A_1 &  & &  \cr \vdots &  & {\cal H}_{2h1p} &  \cr A_L &  &  & \cr} \;
.\label{eq:bag3}
\ee
In this way, the BAGEL(N,M) approximation reduces the matrix defined in
eq.(\ref{eq:matr1}) to a subspace of dimension $l*(N+M+1)$, if $l$ is
the number of single-particle states of the same symmetry. The Green
function of this BAGEL(N,M) approximation is then defined
in the manner discussed in the preceding section but now considering only
the eigenvalues and eigenvectors of the truncated matrix. It is obvious
that the BAGEL(0,0) corresponds to the HF approximation, while for large
enough $N$ and $M$ the BAGEL(N,M) becomes identical to the exact solution
of eq.~(\ref{eq:matr1}).

\section{Results and Discussion}

\subsection{The Spectral Function}

The spectral function summarizes the spectroscopic factors for adding,
at energies above the Fermi energy $E_{F}$, or removing, at energies
below $E_{F}$, nucleons from the nucleus under consideration.
For a model space with discrete single-particle orbits, as we consider
in the present investigation, the spectral function is given in terms
of $\delta$ functions at the various poles of the single-particle Green
function (see eq.~(\ref{eq:dyson2})).  For the sake of presentation, we
will fold these $\delta$ functions with a gaussian distribution
assuming a with $\Gamma$. This width can be interpreted as a simple way
to include the escape width for the states of the residual nuclei, or as
a tool to express the uncertainty of the theoretical calculation.
The second interpretation applies in particular to the use of the BAGEL
approximation. With this folding
procedure the spectral function is given by
\begin{eqnarray}
S_{\alpha}(\omega ) = \frac{1}{ \sqrt{\pi}\Gamma} \Bigl[ & \sum_{n}
\exp{-\frac{(\omega - \omega_{n}^+)^2}{\Gamma^2}} \vert < \Psi_n^{A+1}
\vert a^\dagger_\alpha \vert \Psi_0 >\vert^2\nonumber \\& + \sum_{m}
\exp{-\frac{(\omega - \omega_{m}^-)^2}{\Gamma^2}} \vert < \Psi_m^{A-1}
\vert a_\alpha \vert \Psi_0 >\vert^2 \Bigr] \; . \label{eq:gaus}
\end{eqnarray}
In this equation, as in eq.~(\ref{eq:norm}), the summation over $n$
is restricted to the poles $\omega_{n}^+$ above $E_{F}$ and the summation
over $m$ to those below the Fermi energy.

As a first example, we present in Fig.~2  results for the spectral
function of the $p_{3/2}$ orbits in \caf\ for energies below the Fermi
energy. In this figure we
compare the results obtained in the BAGEL(M=10, N=10) approach to
those obtained using  M=N=20. If we choose for the
presentation a width of $\Gamma$ = 2 MeV (lower part of the figure)
differences between the two approximations are clearly visible. It is
not only that the M=N=20 distribution displays more poles, as the
corresponding Green function contains more poles, but also the position
of the peaks are shifted if the accuracy of the BAGEL approximation is
improved. If, however, we examine the same results with a reduced
energy resolution, obtained  by increasing the width
in eq.(\ref{eq:gaus}) to $\Gamma$=5 MeV, the differences between the 2
BAGEL approximation become invisible. This is a demonstration of the
fact that the BAGEL(M,M) approximation reproduces all energy-weighted
moments of the exact calculated distribution from order $n=0$ to the
order $n=2M+1$ \cite{sko2}. In the following discussion, we will
present results for the BAGEL(20,20) approximation and display the data
assuming a width $\Gamma$=2MeV.

The spectral functions displayed in figures 3 and 4 have been obtained
by multiplying the function $S_{\alpha}$ above with the degeneracy
factors $2j_{\alpha}+1$ and adding up the various contributions with
the same orbital angular momentum. Fig.~3  shows the spectral functions
for \osi\ in the energy interval ranging from $\omega$ = -50
MeV to 30 MeV. The results in this figure clearly  reflect the shell-structure
of nuclei and the limitations of the independent particle model (IPM).
For orbital angular momentum $l=0$
one observes a strong peak at -4 MeV, just above the Fermi energy,
which can be identified with the $1s_{1/2}$ state of the IPM. The
single-particle strength for the $0s_{1/2}$ and $2s_{1/2}$ states is
distributed in energy intervals around -40 MeV and around +20 MeV,
respectively.

Similar features are also observed for the $l=1$ states. The strength
below the Fermi energy is mainly located at an energy $\omega$ around
-18 MeV. This peak originates from two poles: one at $\omega$=-19.12
MeV ($p_{3/2}$), the other at -16.73 MeV ($p_{1/2}$). Each of these
poles carries around 75\% of the total strength below the Fermi energy
in $p_{3/2}$ and $p_{1/2}$ states, respectively. With the energy
resolution ($\Gamma$ = 2 MeV) considered for the figures, the
spin-orbit splitting is not resolved. It is worth noting that the
contributions to the self-energy beyond HF reduce the calculated
spin-orbit splitting for the $l=1$ hole states from 3.6 MeV to 2.4 MeV.
This last number is calculated by taking the energy difference between the two
states carrying maximum strength. The decrease in the spin-orbit splitting
is in agreement with the results of previous calculations \cite{zamick},
which suggest that relativistic effects are mainly responsible
for the spin-orbit splitting between hole states.

Additional $l=1$ single-particle strength is observed at lower energies,
although the individual contributions are so small that they are barely
visible in Fig.~3. It is worth noting that the total occupation probability
(see eq.~(\ref{eq:occu})) is 0.926 (0.901) for the $p_{3/2}$ ($p_{1/2}$)
state while the contribution of the main peak to this value is only 0.781
(0.790). The additional strength  originates from states at lower energies,
which implies larger excitation energies in the A-1 nucleus.
The $l=1$ spectral function furthermore displays the
single-particle strength at positive energies, which corresponds to the
$1p$ and $2p$ states of the IPM.

In the top part of Fig.~3 the spectral function for the $l=2$ and $l=3$
states are displayed. In the IPM these states are unoccupied. Due to
the correlations included in our approximation, we observe an
occupation probability of 0.052 and 0.010 for the $d$ and $f$-states,
respectively. In the case of the $l=2$ spectral function one can
observe two peaks originating from the $d_{5/2}$ and $d_{3/2}$ states.
The calculated spin orbit splitting between these states is 4.92 MeV,
slightly below the HF result of 5.41 MeV. For these particle states,
the result with and without correlations, is in good agreement with the
experimental value of 5.09 MeV. This is again in line with the
arguments given in ref.~\cite{zamick}, as the relativistic effects
discussed there should not influence the spin-orbit splitting for the
particle states.

In Fig.~4 we show the spectral function calculated for \caf ,
considering the orbital angular momenta $l=0$ to $l=3$. The
presentation is limited to energies below the Fermi energy, since for
these energies a comparison with experimental data, deduced from
(e,e')p experiments is possible \cite{kramer}. In the case of $l=0$,
the spectral strength for the hole-state close to the Fermi energy is
nicely reproduced. The broad distribution of $l=0$ strength originating
from deep-lying hole states, which is predicted in the calculation down
to energies of -100 MeV, has not been resolved in the experimental
data. Also for \caf\ the spin-orbit splitting calculated for the $d$
hole states is smaller than the observed splitting. In the
experimental data one finds a spectral strength below the Fermi-energy
for $l=3$ (notice the modified scale in the lower right part of Fig.~4).
Such a strength is completely absent in the IPM or HF approach.
Including correlation effects we observe a strength of similar size as
the experimental data. While, however, the experimental strength is
located in essentially one state, the calculation yields a distribution
over various states. The agreement between theory and experiment is
reasonable, keeping in mind that the theoretical results have been
derived from a realistic NN interaction without any adjustable
parameter. Results on the occupation probabilities obtained for \caf\
considering various approximations for the nucleon self-energy are
listed in table \ref{ta:occu}.

\subsection{Correlations and Ground-State Properties}

A major aim of the present investigation is to study the effects of
correlations included in the definition of the nucleon self-energy on
the bulk properties of closed shell nuclei. For that purpose we
consider various approximations. The first of these is what we will call the
Hartree-Fock (HF) approximation, i.e.~the self-energy is calculated in
the HF approximation. Note, however, that this HF approximation is
something in between a conventional HF and BHF calculation. The
G-matrix interaction (see section 2.1) accounts for intermediate
2-particle states, which are outside the considered model space.
However, in contrast to a self-consistent BHF calculation it does not
account for 2-particle states between the Fermi level and the limit of
the model space. Such correlations are taken into account in the
subsequent approximations.  Furthermore the
hamiltonian has been adjusted in such a way that this HF approximation
leads to a value for the radius of the nuclei \osi\ and \caf , which
are close to the experimental values (eq.(\ref{eq:adjust})).
Therefore, in studying effects of
correlations beyond HF, one should pay attention only to the effects in
the calculated radii relative to the HF result and not expect any
improvement compared to the experiment.

The second approximation considered in our study consists in adding the
contribution of the second-order (2p1h) diagram (a) in Fig.~1 to the
self-energy. This approximation will be labeled ``bare 2p1h'' approach in the
tables and the discussion below. A part of the residual interaction in the
space of 2p1h configurations is given by the interaction between the
two particle lines. The approximation which accounts for this part of
the residual interaction and includes the diagram of Fig.~1c plus all
other particle-particle ladder diagrams in the definition of the
self-energy, corresponds to the Brueckner - Hartree-Fock approach and
will be identified in the following as "BHF". On the other hand the term
"2p1h" describes the approximation in which the whole residual interaction
is considered in the 2p1h subspace (diagram of Fig.~1d and others).
Finally, the term ``total'' describes the calculation in the complete
space of 2p1h and 2h1p configurations with all residual interactions
included.  This means that the complete matrix in
eq.(\ref{eq:matr1}) extended by the modifications of
eq.(\ref{eq:resid}) is considered to construct the Green function.

For all studies beyond HF the single-particle Green function has been
determined in the BAGEL(M=10,N=10) approximation. The results on the
bulk properties of nuclei discussed below turned out to be extremely
stable with respect to this approximation. Increasing the accuracy of
the BAGEL to M=N=20, did not lead to any change in the results
displayed in the tables.

For the resulting Green functions we have determined the occupation
probabilities $N_{\alpha}$ according to eq.~(\ref{eq:occu}) and a mean
value for the spectral distribution below the Fermi energy, defined by
\be
\epsilon_{<,\alpha} = \frac{1}{N_{\alpha}}
\int_C {d \omega \over 2 \pi i}  \omega G_{\alpha\alpha} (\omega )
=  \frac{1}{N_{\alpha}} \sum_m \omega_{m}^- \vert <\Psi_m^{A-1}
\vert a_\alpha \vert \Psi_0 > \vert^2 \; . \label{eq:mean}
\ee
The contour integral in the first part of the equation is defined in
the same way as in eq.~(\ref{eq:observ1}) and as before the summation in
the second part of the equation is restricted to poles in the Green
function with an energy $\omega_{m}^-$ below the Fermi energy. In the
case of the HF approximation the definition in eq.~(\ref{eq:mean}) leads
to the HF single-particle energy. The modification of these mean
values due to the correlations give some insight into the effects of
correlation on the calculated binding energy (compare
{}~eq.(\ref{eq:energy}).

Results for the occupation probabilities and the $\epsilon_{<}$ are
listed in table \ref{ta:occu} for some single-particle orbits of \caf .
One finds that the admixture of 2p1h configurations reduces the
occupation probability by 3\% for the deep-lying hole states and
by around 5\% for states closer to the Fermi level. Inclusion of
the residual interaction between the 2p1h configurations slightly
enhances these depletions for the occupation probabilities. The effects
of the residual interaction are dominated by the pp ladder diagrams while the
additional interaction terms can essentially be ignored. The removal of
occupation strength from the state below the Fermi energy to 2p1h
configurations above $E_{F}$ is accompanied by a lowering of the mean
values $\epsilon_{<}$ below the HF single-particle $\epsilon^{HF}$.
This is obvious since one has the following relation \cite{sko1}
\be
\sum_{\alpha} N_{\alpha}\epsilon_{<,\alpha} + (1-N_{\alpha})
\epsilon_{>,\alpha} = \sum_{\alpha}
\epsilon_{\alpha}^{HF}\; ,
\label{eq:sumrul}
\ee
where the sum on $\alpha$ is a sum on all single-particle orbits of the
same symmetry and the mean value for the spectral distribution above
the Fermi energy is defined in
analogy to eq.~(\ref{eq:mean}) by
\be
\epsilon_{>,\alpha} = \frac{1}{1-N_{\alpha}}
\sum_n \omega_{n}^+ \vert <\Psi_n^{A+1}
\vert a_\alpha^\dagger \vert \Psi_0 > \vert^2 \; . \label{eq:meanp}
\ee
It is this lowering of the mean values $\epsilon_{<}$, which is the
main source for the gain in energy per particle when we evaluate the
energy according to eq.~(\ref{eq:energy}) using the ``bare 2p1h'', ``BHF''
and ``2p1h'' approximations.  The calculated values for the energy per particle
for \osi\ and \caf\  are listed in table \ref{ta:ca40}. As seen in this table,
the inclusion of 2p1h configurations within our model space increases the
calculated binding energies by 1.9 MeV and 1.6 MeV per nucleon for \osi\ and
\caf , respectively. As already discussed above, the effects of the residual
interaction between 2p1h configurations, are small (0.3 MeV per
nucleon) and are mainly due to the particle-particle interactions
contained in the BHF approach. The effect of the 2p1h admixture on the
calculated radii is essentially negligible.

It should be mentioned that the way we calculate the energy per nucleon
in the BHF approximation is slightly different from the conventional
form. In conventional BHF calculations one determines the total energy
from the mean values $\epsilon_{<}$, which, for the corresponding
approximation in the self-energy, are identical to the BHF,
$\epsilon^{BHF}$, single-particle energies and calculates the total
energy from
\be
E_{conv}^{BHF} = \frac{1}{2} \sum_{\alpha} \left[ <\alpha \vert T \vert
\alpha > + \epsilon_{\alpha}^{BHF} \right] n_{\alpha}\; ,
\label{eq:bhfe}
\ee
with $n_{\alpha}$ the occupation probabilities of the HF or IPM
approach. This expression essentially corresponds to
eq.~(\ref{eq:energy}) except that the conventional BHF approach ignores
the differences in the occupation probabilities for the various states.
The calculated binding energy per nucleon according to
eq.~(\ref{eq:energy}), using for the particle number the results of
eq.~(\ref{eq:numb})  are very close to those obtained in the
conventional way (differences are typically around 0.1 MeV).

The inclusion of 2h1p configurations in the ``total'' calculation leads
to a fragmentation of the strength into many states below the Fermi
energy. One finds, that occupation probabilities and mean
values for the spectral distribution $\epsilon_{<}$ are not very much
affected by the inclusion of these additional correlations (see table
\ref{ta:occu}) for the states which are occupied in the IPM. The total
approach, however, leads to a non-vanishing occupation also for
configurations with quantum numbers $\alpha$, which are unoccupied in
the IPM. The mean value $\epsilon_{<}$ for these 2h1p configurations is
typically quite attractive. Therefore one observes a non-negligible
gain in the binding energy due to the hole-hole correlations. In our
standard model space this gain is 0.4 MeV and 0.3 MeV per nucleon for
\osi\ and \caf , respectively.

Considering just the energies it is obvious, that the effect of 2p1h
correlations is significantly larger than those obtained from 2h1p.
This would support the BHF approximation, which includes the effects of
the former kind but ignores 2h1p contributions. For the calculation of
the nuclear radii, however, the 2h1p correlations seem to be more
important than those deduced from particle-particle correlations. The
inclusion of 2h1p correlations yields an increase for the calculated
radii of around 0.1 fm. Alone this increase does not seem to be
dramatic. It is worth noting, however, that the combination, an increase of
calculated binding energy and radius, moves the result for these bulk
properties of nuclei away from the so-called ``Coester band'' \cite{carlo}.
This is a similar feature as has also been observed for hole-hole
correlations in infinite nuclear matter \cite{kuo1,kuo2}.

The sensitivity of our results on the choice of the model space and on
the parameters of the residual interaction is examined in table \ref{ta:o16}.
This table contains results for \osi\ for three choices of the model space:
one with 10 active orbits, another with 15 and a third where 21 active
orbits are considered. This table contains results for the energies per
nucleon obtained in the HF approximation, for the ``total'' approach
and the difference between these 2 values. Since the effective
interaction depends on the choice of the model space, also the HF
results change significantly. For the small model space ($N_{Cut}$=10)
the effects of 2p1h correlation with 2 particles in sdg-shell or above
are contained already in the interaction. Therefore a larger binding
energy is obtained in the HF approximation for this small model space
than for the HF calculation using the interaction which is appropriate
for a larger model space.

These effects of the 2p1h correlations and the interference with 2h1p
correlations are contained in the ``total'' calculation. The
convergence of these correlation effects is rather slow with increasing
model space. Both the calculated binding energies as well as the
results for the radii are enlarged if the number of shells included in
the model space is increased. Therefore the features discussed above
may get enhanced if even larger model spaces are considered.

Table \ref{ta:o16} displays results also for an alternative choice
for the starting energy in eq.~(\ref{eq:betheg}), using $\omega = -5 MeV$
instead of our standard choice $\omega = -30 MeV$. One observes that
the larger value for $\omega$ yields larger energies already in the HF
approximation and also for the ``total'' calculation. The correlation
effects reflected in
the energy differences between these two approximations are rather
insensitive to the choice of $\omega$.

\section{Conclusions}

The effects of correlations beyond the Hartree-Fock approximation are
investigated in large model spaces including up to 21 active
single-particle states of the harmonic oscillator potential. The
residual interaction for this model space is derived from a realistic
meson-exchange potential \cite{rupr}, accounting for short-range correlation
effects beyond the model space by solving an appropriate
Bethe-Goldstone equation. The correlations within the model space are
evaluated in terms of the single-particle Green function. The Green
function can easily be evaluated even for a large model space by
employing the BAGEL \cite{sko1,sko2} approximation.

The spectral functions evaluated from this Green function for \osi\ and
\caf\ yield a
distribution of single-particle strength over a broad range of
energies.
A comparison with an empirical spectral function for \caf ,
determined from $(e,e')p$  experiments \cite{kramer} demonstrates
that the spectral
function, calculated without adjustable parameter, produces a reasonable
agreement. However, the spin-orbit splitting between hole states close to the
Fermi energy is underestimated even after the inclusion of correlation
effects. It is argued that relativistic features, not considered in the
present work, are mainly responsible for this splitting \cite{zamick}.

Special attention is paid to the effects of correlation, treated in
various approximations, on the calculation of bulk properties of
nuclei,
like binding energy and radius. It turns out that the effects of
correlation on the calculated energies are dominated by the
particle-particle (2p1h) correlations taken into account in the BHF
approximation. Some additional binding energy is obtained if also
hole-hole (2h1p) correlations are considered. The 2h1p correlations
furthermore give rise to an increase for the calculated radius.
Therefore the properties evaluated for ground states of nuclei are
``moved off the Coester band'' towards the experimental data \cite{carlo}.
This effect of hole-hole correlations is similar to the one observed in
nuclear matter \cite{kuo1,kuo2}.

The effects obtained from inclusion of the hole-hole correlations are
not  very large. In the model space under investigation they yield an
extra gain in binding energy of around 0.4 MeV per nucleon and an
increase of the radii of around 0.1 fm. The effects may get larger, if
larger model spaces are taken into account. This is possible with the
techniques explored here. However, for a realistic investigation one
should try to represent the particle states in terms of plane waves
rather than harmonic oscillator states.

This work has partly been supported by the Deutsche Forschungsgemeinschaft
(DFG, Fa67/14-1).

\clearpage
\begin{table}
\caption{Occupation probabilities and mean values for the energy
distribution below the Fermi energy, calculated for different
single-particle states in \caf . The different columns correspond to
various approximations used in the calculation for the self-energy of the
nucleons.
The column labeled ``HF'' stands for the Hartree-Fock approximation, whilein
the
second column also the second-order diagram of Fig.1a has been taken
into account (``bare 2p1h''). All pp ladder diagrams (like e.g. the
diagram of Fig.1c) are included in the Brueckner-HF (``BHF'') approximation.
If all residual interactions between 2p1h configurations (also Fig.1d)
are taken into account, one
obtains column 4 (``2p1h''). The results of the complete calculation,
also including the 2h1p configurations with residual interactions are
given in column 5 (``total''). The single-particle energies are given
in MeV.}
\label{ta:occu}
\begin{center}
\begin{tabular}{||r|rrrrr||}
\hline\hline
&&&&&\\
&\multicolumn{1}{c}{HF}&\multicolumn{1}{c}{bare 2p1h}&
\multicolumn{1}{c}{BHF}&\multicolumn{1}{c}{2p1h}&
total \\
&&&&&\\ \hline
&&&&&\\
$0s_{1/2}$, $\epsilon_{<}$ & -58.211 & -61.098 & -61.189 & -61.126 &
-61.368 \\
$N_{\alpha}$ & 1.0000 & 0.9722 & 0.9704 & 0.9714 & 0.9689 \\
&&&&&\\
$0p_{3/2}$, $\epsilon_{<}$ & -38.796 & -41.552 & -41.695 & -41.649 &
-41.876 \\
$N_{\alpha}$ & 1.0000 & 0.9672 & 0.9637 & 0.9642 & 0.9610 \\
&&&&&\\
$0p_{1/2}$, $\epsilon_{<}$ & -35.901 & -39.074 & -39.236 & -39.165 &
-39.432 \\
$N_{\alpha}$ & 1.0000 & 0.9618 & 0.9577 & 0.9589 & 0.9553 \\
&&&&&\\
$0d_{5/2}$, $\epsilon_{<}$ & -20.747 & -23.597 & -23.905 & -23.941 &
-24.195 \\
$N_{\alpha}$ & 1.0000 & 0.9511 & 0.9383 & 0.9335 & 0.9267 \\
&&&&&\\
$0d_{3/2}$, $\epsilon_{<}$ & -19.446 & -19.861 & -20.231 & -20.170 &
-20.509 \\
$N_{\alpha}$ & 1.0000 & 0.9363 & 0.9206 & 0.9199 & 0.9112 \\
&&&&&\\
$1s_{1/2}$, $\epsilon_{<}$ & -16.350 & -22.661 & -23.003 & -23.009 &
-23.281 \\
$N_{\alpha}$ & 1.0000 & 0.9439 & 0.9298 & 0.9259 & 0.9187 \\
&&&&&\\
$0f_{7/2}$, $\epsilon_{<}$ &         &         &         &         &
-40.058 \\
$N_{\alpha}$ & 0.0    & 0.0    & 0.0    & 0.0    & 0.0459 \\
&&&&&\\
$0f_{5/2}$, $\epsilon_{<}$ &         &         &         &         &
-42.952 \\
$N_{\alpha}$ & 0.0    & 0.0    & 0.0    & 0.0    & 0.0444 \\
&&&&&\\
$1p_{3/2}$, $\epsilon_{<}$ &         &         &         &         &
-48.003 \\
$N_{\alpha}$ & 0.0    & 0.0    & 0.0    & 0.0    & 0.0381 \\
&&&&&\\
$1p_{1/2}$, $\epsilon_{<}$ &         &         &         &         &
-48.254 \\
$N_{\alpha}$ & 0.0    & 0.0    & 0.0    & 0.0    & 0.0391 \\
&&&&&\\
\hline\hline
\end{tabular}
\end{center}
\end{table}
\clearpage
\begin{table}
\caption{Results for binding energy per nucleon (E/A), radius of the nucleon
distribution and particle number A obtained in various approximations
(see caption of table 1). The effects of the Coulomb energy and the
charge radius of the proton have been removed from the experimental
data to allow a direct comparison to the calculated values. Results are
given for \osi\ and \caf . For the calculation of \osi\ particle states
up to and including the sdg shell are taken into account, while in the
case of \caf\ all states up to the hfp shell are considered. All
energies are listed in MeV and the radii in fm.}
\label{ta:ca40}
\begin{center}
\begin{tabular}{||c|rrr|rrr||}
\hline\hline
&&&&&&\\
&\multicolumn{3}{c|}{\osi}&\multicolumn{3}{c||}{\caf} \\
&&&&&&\\
&\multicolumn{1}{c}{E/A}&\multicolumn{1}{c}{Radius}&\multicolumn{1}{c|}{A}
&\multicolumn{1}{c}{E/A}&\multicolumn{1}{c}{Radius}&\multicolumn{1}{c||}{A}
 \\
&&&&&&\\ \hline
&&&&&&\\
HF & -4.202 & 2.633 & 16.000 & -6.491 & 3.455 & 40.000 \\
bare 2p1h & -6.099 & 2.626 & 15.159 & -8.104 & 3.451 & 38.152 \\
BHF & -6.373 & 2.625 & 14.894 & -8.275 & 3.449 & 37.765 \\
2p1h & -6.370 & 2.625 & 14.868 & -8.278 & 3.448 & 37.700 \\
total & -6.747 & 2.720 & 16.262 & -8.587 & 3.533 & 40.379 \\
&&&&&&\\ \hline
&&&&&&\\
Exp & -9.118 & 2.580 & 16.000 & -10.597 & 3.410 & 40.000 \\
&&&&&&\\ \hline\hline
\end{tabular}
\end{center}
\end{table}
\clearpage
\begin{table}
\caption{Results for the energy per nucleon in the HF approximation, in the
``total'' approximation and the gain in energy $\Delta E$ relative to HF,
obtained in various model spaces for the nucleus \osi .
Furthermore the
radius of the nucleon distribution and the particle number A are listed
calculated in the ``total'' approximation (see tables 1 and 2). The
various model spaces are identified in terms of $N_{Cut}$ the number of
active single-particle orbits.}
\label{ta:o16}
\begin{center}
\begin{tabular}{||c|rrrrr||}
\hline\hline
&&&&&\\
$N_{Cut}$&\multicolumn{1}{c}{$E_{HF}/A$}&\multicolumn{1}{c}{$E_{tot}/A$}
&\multicolumn{1}{c}{$\Delta
E/A$}&\multicolumn{1}{c}{Radius}&\multicolumn{1}{c||}{A}
 \\
&&&&&\\ \hline
&&&&&\\
&\multicolumn{5}{c||}{$\omega$=-30 MeV} \\
&&&&&\\
10 & -4.791 & -6.388 & -1.597 & 2.683 & 16.155 \\
15 & -4.202 & -6.747 & -2.545 & 2.720 & 16.262 \\
21 & -3.628 & -7.052 & -3.424 & 2.754 & 16.379 \\
&&&&&\\ \hline
&&&&&\\
&\multicolumn{5}{c||}{$\omega$=-5 MeV} \\
&&&&&\\
10 & -6.455 & -8.171 & -1.716 & 2.686 & 16.162 \\
15 & -5.556 & -8.214 & -2.658 & 2.723 & 16.279 \\
21 & -4.863 & -8.398 & -3.535 & 2.756 & 16.384 \\
&&&&&\\ \hline\hline
\end{tabular}
\end{center}
\end{table}
\clearpage
\section{Figure Captions}
{\bf Figure 1:}
{Diagrams representing various contributions to the self-energy
of second and higher order in the residual interaction ${\cal G}$.}

\bigskip\bigskip
{\bf Figure 2:}
{The spectral function defined in  eq.~(29) obtained for the
$p_{3/2}$ states in \caf , employing two different BAGEL
approximations. The curves in the upper part are obtained assuming a
width of $\Gamma$=5 MeV, whereas in the presentation displayed in the
lower part $\Gamma$=2 MeV has been used.}

\bigskip\bigskip
{\bf Figure 3:}
{Spectral functions for the nucleus \osi\ in a region of
energies between $\omega$=-50 MeV to $\omega$= 30 MeV. The spectral
functions for the various orbital angular momenta ($l=0,1,2,3$) have
been obtained from eq.(29) (width $\Gamma$=2 MeV) by multiplying the
contributions from the different total angular momenta
$j_{\alpha}=l_{\alpha}\pm 1/2$ by the degeneracy factors
$2j_{\alpha}+1$.}

\bigskip\bigskip
{\bf Figure 4:}
{Spectral functions obtained for the nucleus \caf\ for the
various orbital angular momenta$l$ in an energy interval [-50 MeV,
-10 MeV]. The experimental data are from ref.~[17]. Note the different
scales for the spectral function in the various parts of the figure}

\end{document}